\documentclass[journal=nalefd, manuscript=article, layout=twocolumn]{achemso}
\usepackage{amsmath}
\usepackage{amssymb}
\usepackage{graphicx}
\usepackage{balance}

\author{E. Pogorelov}
\email{evgeny.pogorelov@uni-bayreuth.de}
\author{M. Fleck}
\author{H. Federmann}
\author{J. Kundin}
\author{H. Emmerich}
\affiliation{Materials and Process Simulation, University of Bayreuth,
Germany}

\title{Spinodal Gap Dependence on Size and Boundary Reaction Rate for Intercalation in Nanoparticles}

\begin{document}

\begin{abstract}
Using mathematical model for intercalation dynamics in phase-separating materials proposed in (Singh, G. K.; Ceder, G.; Bazant, M. Z.
Electrochimica Acta 2008, 53, 7599) and developed further in (Burch, D.; Bazant, M. Z. Nano Letters 2009, 9, 3795) we found that the spinodal gap generally shrinks as the particle size decreases. We also got that for some range of boundary reaction rate parameter and particle size the concentration spinodal gap is not continuous but it has stable ``islands'' where no spinodal decomposition is expected. In fact the presence of infinitesimally small boundary reaction rate parameter will destabilize nano particle even for infinitesimal length of particle. But then further continuous raise of this parameter will stabilize the system  till some limit. We made careful analysis to study the dependence of spinodal decomposition on boundary reaction rate and size parameters which can be used for LiFePO4 battery material application. Our model predicted that for nanoparticles having size less than diffusion length and enough small boundary reaction rate spontanious charge or discharge will occur.
\end{abstract}

Intercalation phenomena on nanoscale level is observed in many systems such as in intercalation compounds of graphite \cite{DresselhausMS2002}, Li-ion battery cathodes\cite{BurchD2009},  DNA\cite{LermanLS1961, RichardsAN2007}, intercalation of potassium into graphite\cite{HollemanAF2001}, iron oxychloride\cite{KikkawaS1983}.

We are focused on Li intercalation in (Li)FeO4 material for Li-ion battery applications. The decomposition phenomena in this case will decrease the efficincy of such batteries preventing full charge/discharge cycle. Nowdays the Li-ion battery technology is developed in two main directions. (i) Decreasing the size of cathode particles to nanoscale prevents to some level spontanious decomposition and improves the power density\cite{YamadaA2001, HuangH2001}. (ii) The use of various nano or micro coating changes the boundary reaction rate with obvious influence on charging rate\cite{RavetNCY2001, HerlePS2004, KangB2009, WangJ2012}.

In recent years the effective models were developed to describe the phase separation in Li-ion materials based on the classical Cahn-Hillard equation. The model developed by Singh et.al. \cite{SinghGK2008} introduces the insertion reaction condition on the crystal boundaries that allows to explain the dynamics of intercalation waves which were observed in experiments. Further it was also shown that the large insertion rate can stabilize nano particles  by suppresion of the phase separation \cite{BurchBazant2009, BaiP2011}.

In this paper we make detailed analysis the influence of boundary reaction rate on the size and structure of spinodal concentration gap based on the model ot the intercolation dynamic. The theory predicts that the presence of infinitesimally small influx parameter will destabilize nano particle even for infinitesimal length of particle. But then further continuous raise of this parameter will stabilize the system  till some limit. In order to prove the result for founded initial concentrations we substitute the noise into linearised  partial differential equation with correspondent boundary conditions. Then we got the same relative growth rates for noises as it was predicted with error within 0.1\%.

{\bf The bulk equations.} Here we rewrite basic equations of Ref. \cite{BurchBazant2009, SinghGK2008} for one dimensional space to make our comment self consistent and more clear. The Cahn-Hillard type chemical potential \cite{CahnHillert1958} of the system is
\begin{equation}
\mu= \frac{\partial g_\text{hom}(c)}{\partial c}-K\frac{\partial^{2}c}{\partial x^{2}},
\end{equation}
where
\begin{equation}
g_\text{hom}(c)=ac(1-c)+k_BT\bigl[c\log c+(1-c)\log(1-c)\bigr].
\end{equation}
In Ref. \cite{BurchBazant2009} the flux is defined as
\begin{equation}
J=-\rho cM\frac{\partial\mu}{\partial x},
\end{equation}
which makes the behaviour of the system asymmetric. We would like to mention that some authors prefer to use symmetric model \cite{TangBelakDorr032011}.
The time evolution of the concentration is
\begin{equation}
\frac{\partial (\rho c)}{\partial t} +\frac{\partial J}{\partial x}=0.
\end{equation}
Employing the above equations we obtain the forth order equation for the concentration
\begin{equation}
\frac{\partial c}{\partial t}=\frac{\partial}{\partial x}\left(cM\frac{\partial}{\partial x}\left(\frac{\partial g_\text{hom}}{\partial c}-K\frac{\partial^2 c}{\partial x^2}\right)\right).
\end{equation}

{\bf The boundary conditions.} The boundary conditions on the left and on the right side are the following
\begin{align}
\frac{\partial c}{\partial x}\biggm|_{x=0,L}&=0\\
(J-\rho_s R)|_{x=0}&=0,\;(J+\rho_s R)|_{x=L}=0,
\end{align}
where $R$ depends on the difference between boundary value of $\mu$ and "external" constant $\mu_e$ of medium outside of nano particle, and $R$ is proportional to influx rate parameter $R_\text{ins}$
\begin{equation}
R = R_\text{ins}\left(1-\exp\left(\frac{\mu-\mu_{e}}{k_{B}T}\right)\right).
\end{equation}

{\bf Linear stability analysis.} For every fixed particle size $L$ we want to find such set of initial concentrations, where infinitesimally small noise will destabilize the system. The difference between maximal and minimal concentration in this set is the spinodal gap.

To linearize the equations (1-8) one should represent
\begin{equation}
c(x,t) =c_0+\epsilon c_1(x,t)+\mathfrak{o}\left(\epsilon^{2}\right),
\end{equation}
where initial concentration $c_0$ satisfies these equations. Therefore we have on the left and right boundaries $\mu_0=\partial g_\text{hom}(c_0)/\partial c=\mu_e$. After linearisation we have
\begin{align}
&\mu_1(x,t)=\frac{\partial^2 g_\text{hom}(c_0)}{\partial c^2}c_1-K\frac{\partial^2 c_1}{\partial x^2},\\
&J_1=-\rho c_0 M \frac{\partial\mu_1}{\partial x},\;
\frac{\partial(\rho c_1)}{\partial t}=-\frac{\partial J_1}{\partial x},\\
&\frac{\partial c_1}{\partial x}\biggm|_{x=0,L}=0,\\
&\biggl(J_1+\frac{\rho_s R_\text{ins}}{k_BT}\mu_1\biggr)\!\biggm|_{x=0}\!\!\!\!\!\!\!=0,
\biggl(J_1-\frac{\rho_s R_\text{ins}}{k_BT}\mu_1\biggr)\!\biggm|_{x=L}\!\!\!\!\!\!\!=0.
\end{align}

Any continuous real (not complex) noise $c_1(x,t)$ for fixed time can be presented as real Fourier transform in space with coefficients depending on time $t$. Then Laplace transform in time will allow to solve differential equation. To get respective dispersion relation one should substitute $c_1=A\exp(ikx+st)$ or $c_1=A\cos(kx)\exp(st)$ into bulk equation, where $s$ is relative growth rate and $k$ is wave number. After substitution we have
\begin{equation}
s=-c_0Mk^2\biggl(\frac{\partial^2 g_\text{hom}(c_0)}{\partial c^2}+Kk^2\biggr).
\end{equation}
Let us consider only real $k$. Taking into account $K>0$, $M>0$ to make possible linear instability $s>0$ we should require
\begin{align}
&\frac{\partial^2 g_\text{hom}(c_0)}{\partial c^2}<0 \Rightarrow c_\text{min}<c<c_\text{max},\\
&c_{\text{min},\text{max}}=\frac{1}{2}\mp\frac{1}{2}\sqrt{1-\frac{2k_BT}{a}}.
\end{align}
Assuming that the solution of (14) $k^2>0$ and searching for instability $s>0$ we find
\begin{equation}
0<s<s_\text{max}=\frac{c_0 M}{4K} \biggl(\frac{\partial^2 g_\text{hom}(c_0)}{\partial c^2}\biggr)^2.
\end{equation}
In this case we have four different non-zero solutions of (14) $k_1$, $-k_1$, $k_2$, $-k_2$, therefore concentration noise solution for fixed $s>0$ can be written as
\begin{align}
c_1(x,t)&=\bigl(A_1\cos(k_1x)+A_2\sin(k_1x)\\
&+A_3\cos(k_2x)+A_4\sin(k_2x)\bigr)e^{st}.\nonumber
\end{align}
Substituting (18) into boundary conditions we get the system of four linear equations for $A_i$, $i=1\dots 4$. To have non-zero solution the determinant of the system should be equal to zero. It is also very helpful to consider resonance solutions for $s=0$ and $s=s_\text{max}$ to find the borders of some regions consisting of $s>0$.
For $s=0$ we have
\begin{align}
&c_1(x,t)=A_1+A_2x+A_3\cos(kx)+A_4\sin(kx),\\
&k_0=\sqrt{-\frac{1}{K}\frac{\partial^2 g_\text{hom}(c_0)}{\partial c^2}}\,.
\nonumber
\end{align}
Then assuming, that the determinant of system is equal to zero, we get
\begin{equation}
R_\text{ins}\sin(k_0L)=0.
\end{equation}
Analogically $s=s_\text{max}$ we have noise in the form
\begin{align}
&c_1(x,t)=\bigl(A_1\cos(kx)+A_2\sin(kx)+A_3x\cos(kx)\\
&+A_4x\sin(kx)\bigr)e^{s_\text{max}t},\;k_m=\sqrt{-\frac{1}{2K}\frac{\partial^2 g_\text{hom}(c_0)}{\partial c^2}}\,,
\nonumber
\end{align}
which is more complicated, but will allow to get analytical curves for two cases $R_\text{ins}=0$ and $R_\text{ins}\to\infty$
\begin{align}
&\sin^2(k_mL)=0,\;R_\text{ins}=0,\\
&\sin^2(k_mL)=-\frac{\partial^2 g_\text{hom}}{\partial c^2}\frac{L^2}{18K}\,,R_\text{ins}\to\infty.\nonumber
\end{align}
Eq. (22 (a)) provides us infinite number of curves for $s=s_\text{max}$, but Eq. (22 (b)) produces only one curve. With the help of Eqs. (20, 22 (b)) it is easy to prove, that the small stripped instability region shown in Fig. 1 d) will not disappear for $R_\text{ins}\to\infty$.

In calculations we used the following dimensionless parameters  and its values taken from Ref. \cite{BurchBazant2009}
\begin{align}
&\frac{a}{k_BT}=5,\frac{L}{\lambda}\in(0,5),\mathcal{R}=R_\text{ins}\frac{\rho_s\lambda}{\rho D},\bar{s}=\frac{s\lambda^2}{D},\\
&\bar{t}=\frac{tD}{\lambda^2},
\bar{x}=\frac{x}{\lambda}, \bar{k}=k\lambda\nonumber, K=\frac{\lambda^2}{k_BT}, M=\frac{D}{k_BT},
\end{align}
which is enough to fully describe the same dimensionless model. To imagine the real size of considered nano particles one can use suggested in Ref. \cite{BurchBazant2009} $\lambda\approx 4\text{ nm}$.

\begin{figure*}
\begin{tabular}{cc}
a)\includegraphics[scale=0.3]{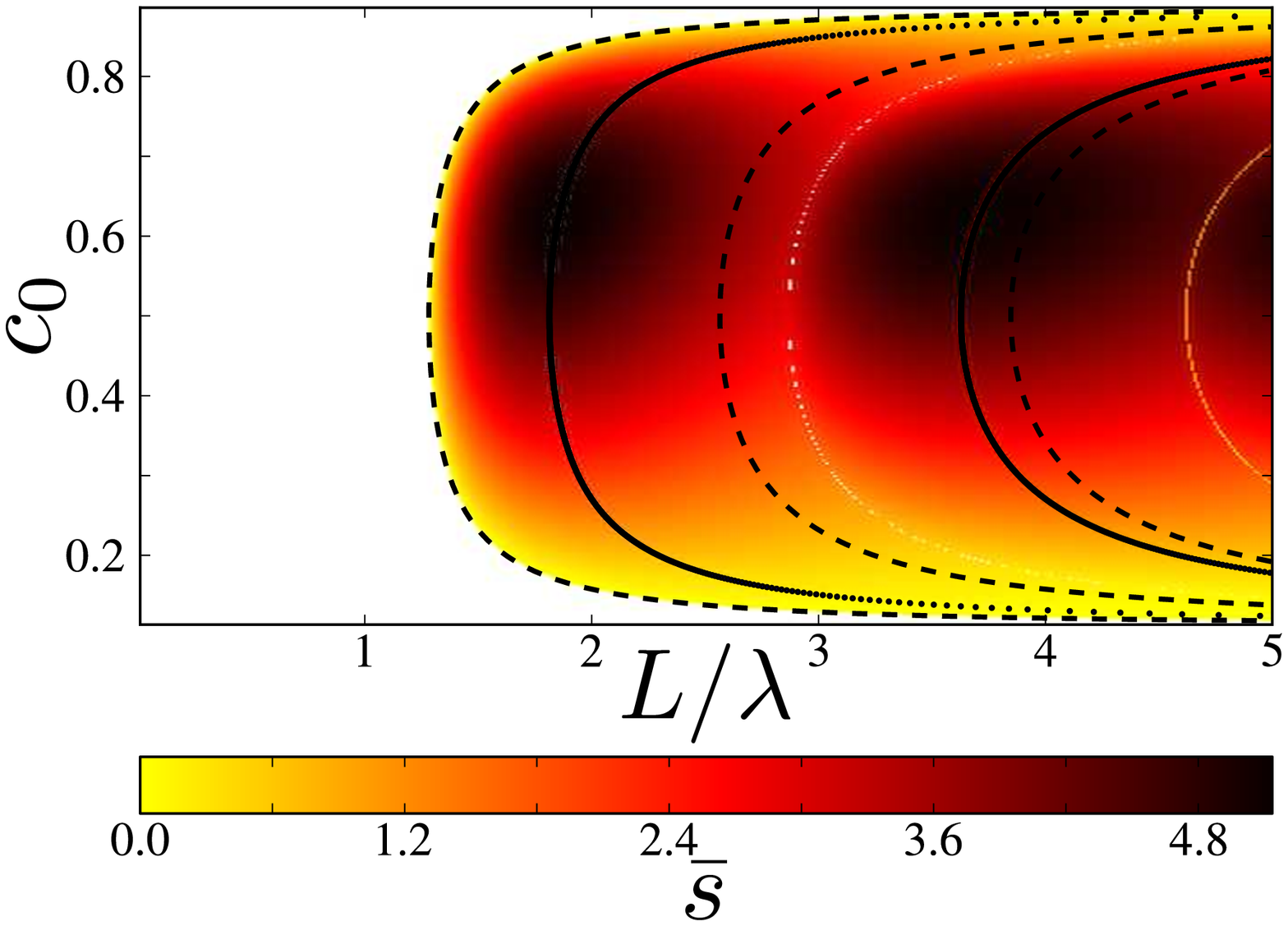}&
b)\includegraphics[scale=0.3]{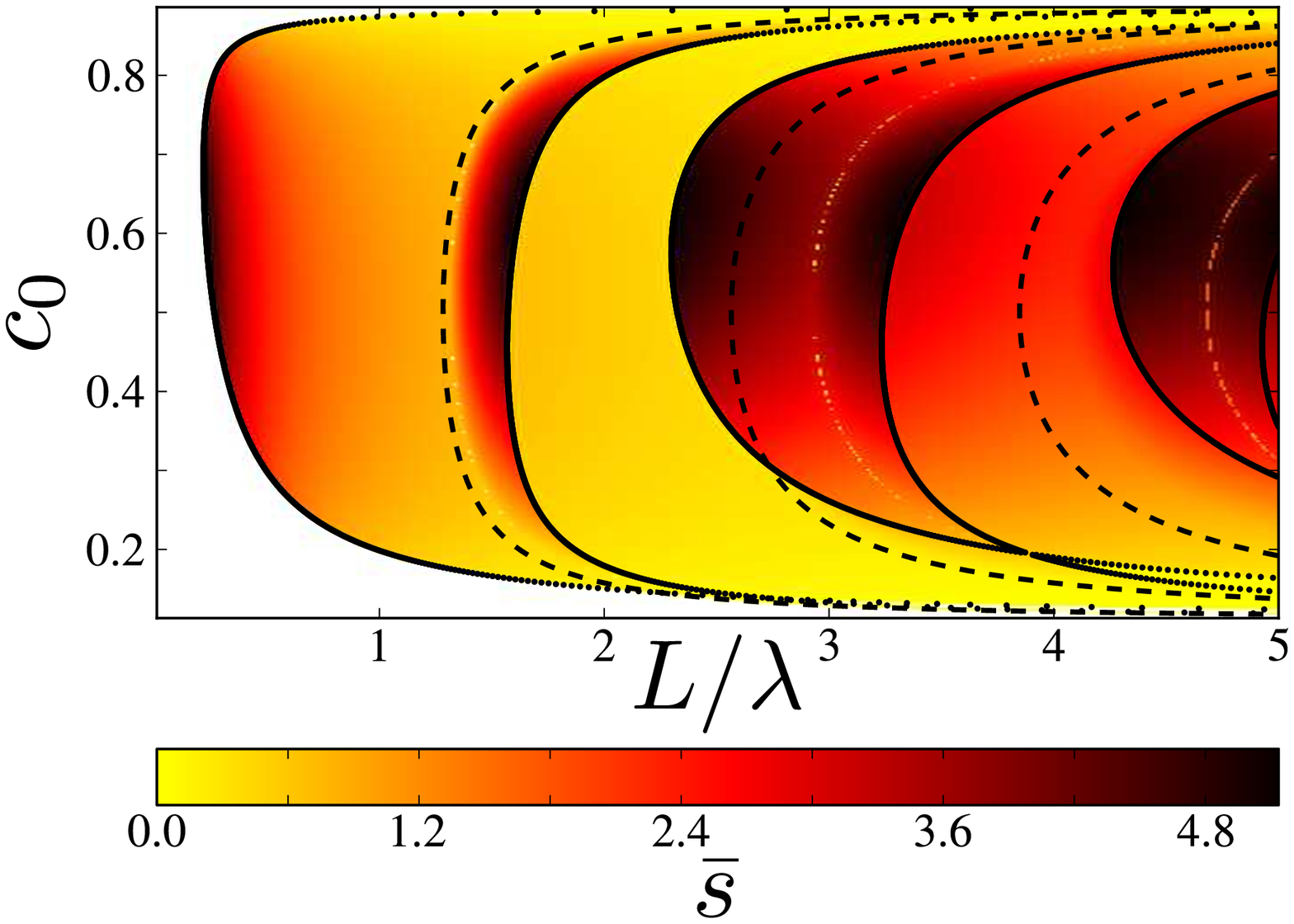}\\
c)\includegraphics[scale=0.3]{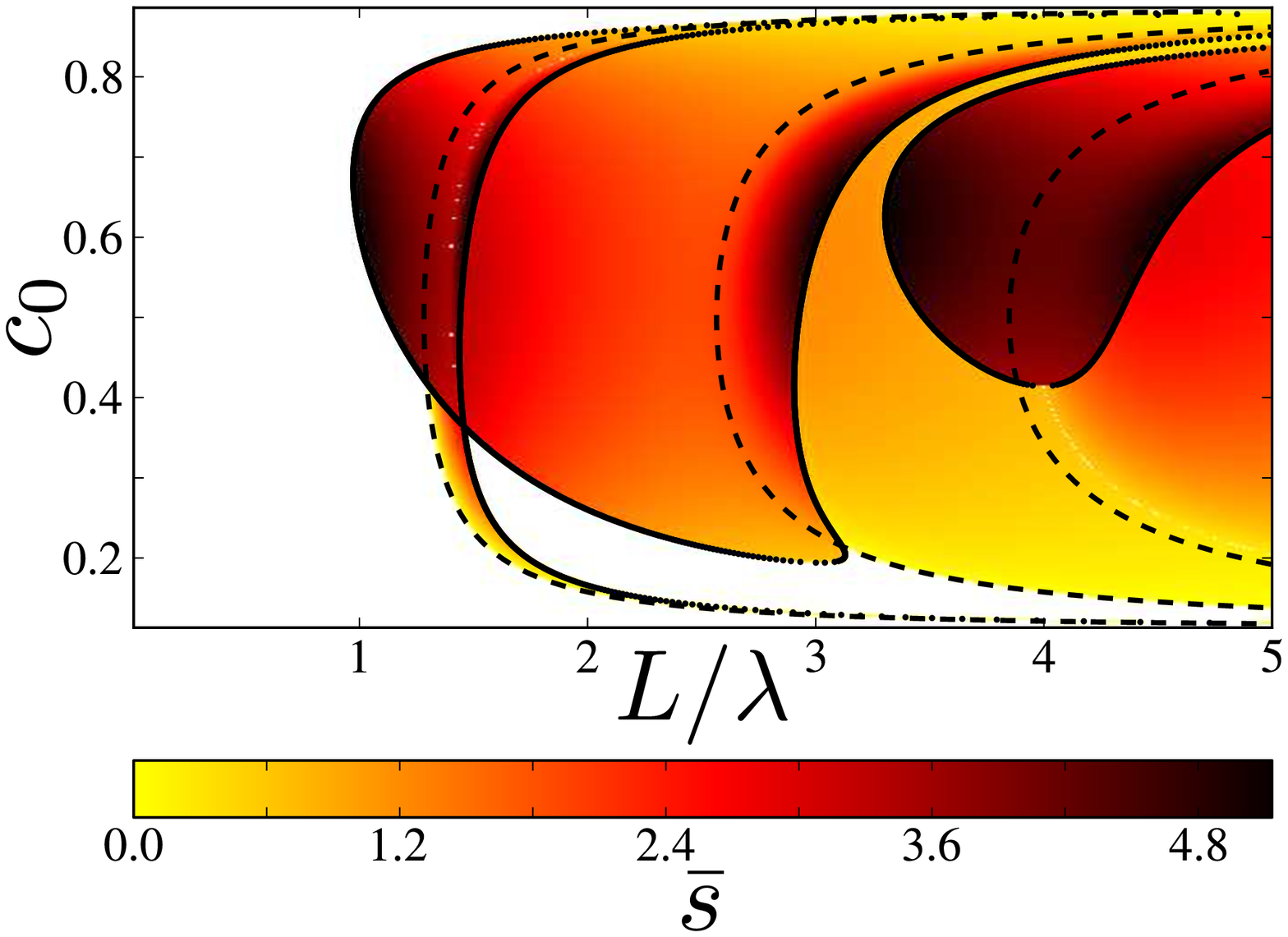}&
d)\includegraphics[scale=0.3]{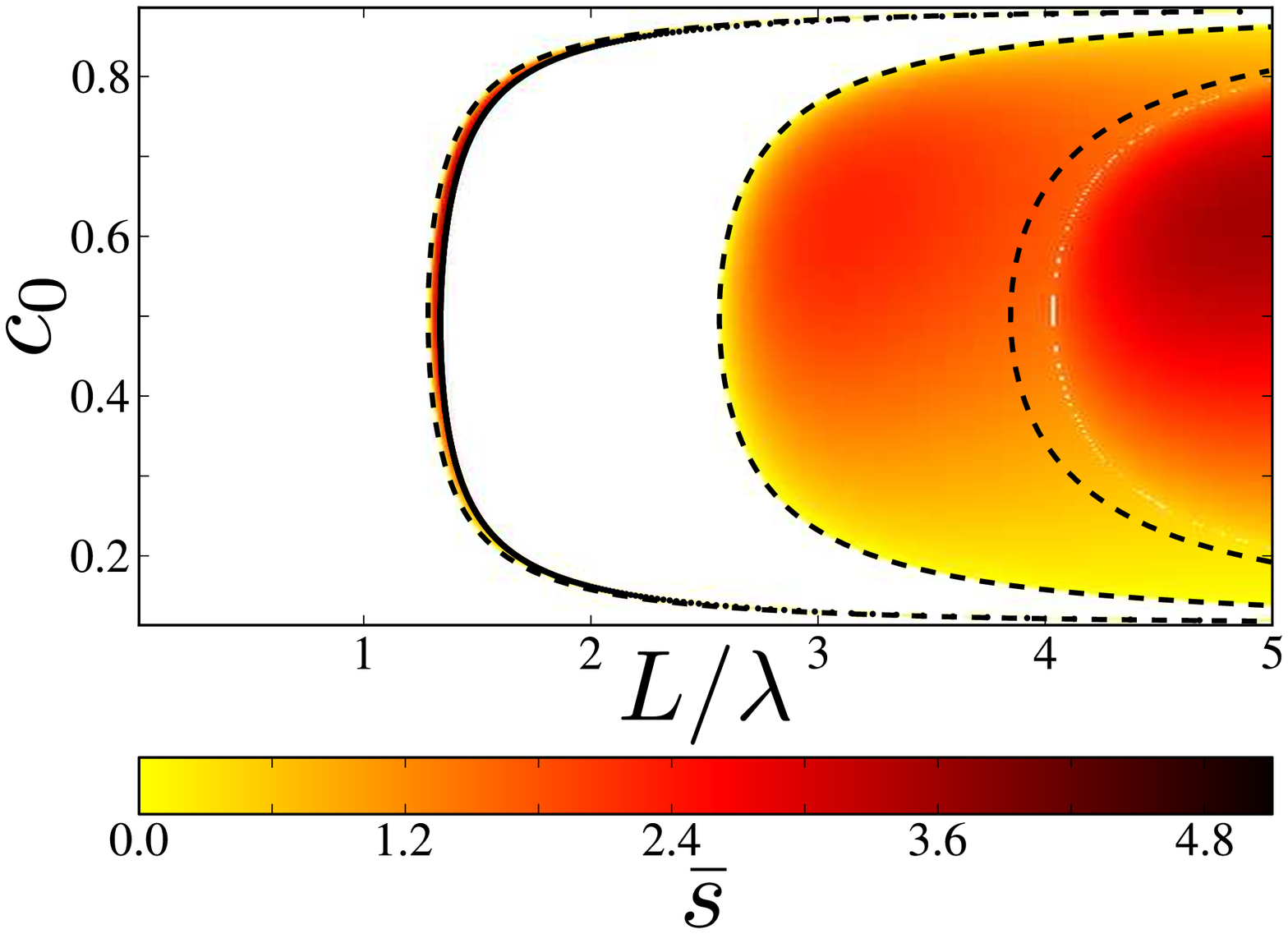}
\end{tabular}
\caption{For vertical axes we have concentration $c_0$, for horizontal axes we have dimensionless particle size $L/\lambda$.
Maximal of existing dimensionless relative growth rate $\bar{s}=s\lambda^2/D$ are shown by color. Dashed lines are related to resonance solution for $\bar{s}=0$. Dotted lines are related to resonance solution for $\bar{s}_\text{max}$. The stable regions are shown by white color. a) $\mathcal{R}=0$; b) $\mathcal{R}=0.1$; c) $\mathcal{R}=0.5$; d) $\mathcal{R}=5$.}
\end{figure*}
In Fig. 1 we show maximal positive relative growth rate $\bar{s}$ by color having solution $c_1(\bar{x},\bar{t})$ which satisfies boundary conditions. In these figures we have initial concentration $c_\text{min}<c_0<c_\text{max}$ as vertical axis and dimensionless length of particle $L/\lambda$ as horizontal axis. Bigger values $\bar{s}>0$ provides greater instability relative growth rate, by white color we notice that there is no solution for $0<\bar{s}<s_\text{max}\lambda^2/D=\bar{s}_\text{max}$. By dashed lines we notice resonance curves $\bar{s}=0$, by dotted lines we notice resonance curve $\bar{s}=\bar{s}_\text{max}$.

\begin{figure*}
\begin{tabular}{c}
a)\includegraphics[scale=0.3]{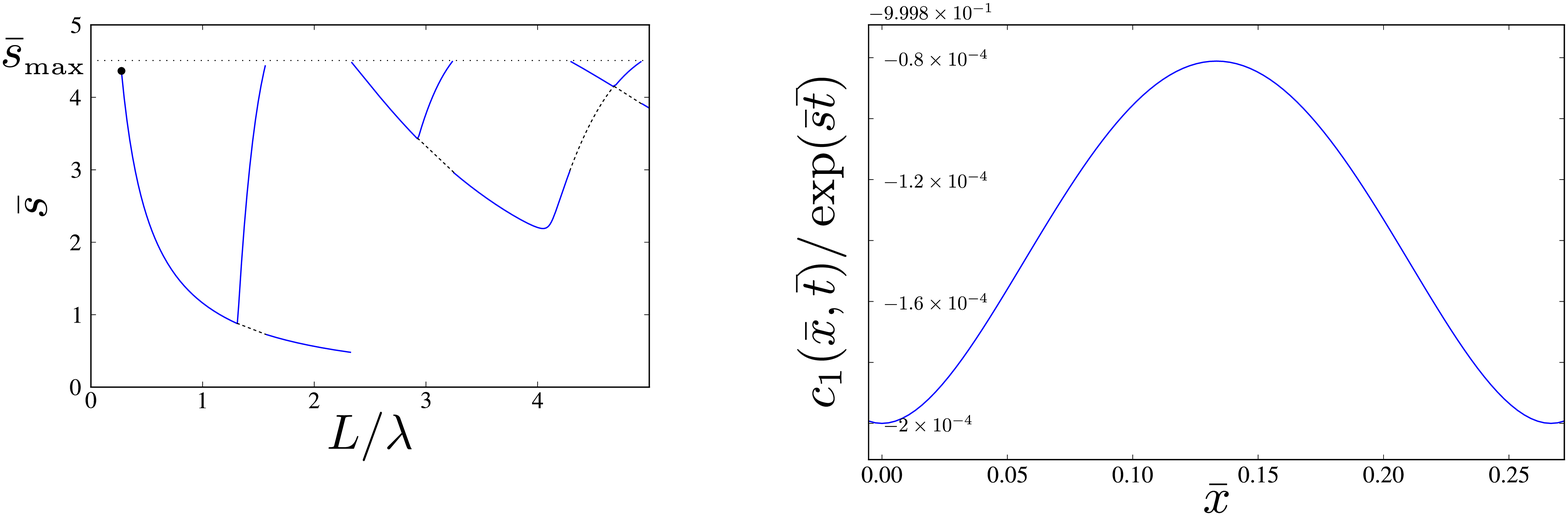}\\
b)\includegraphics[scale=0.34]{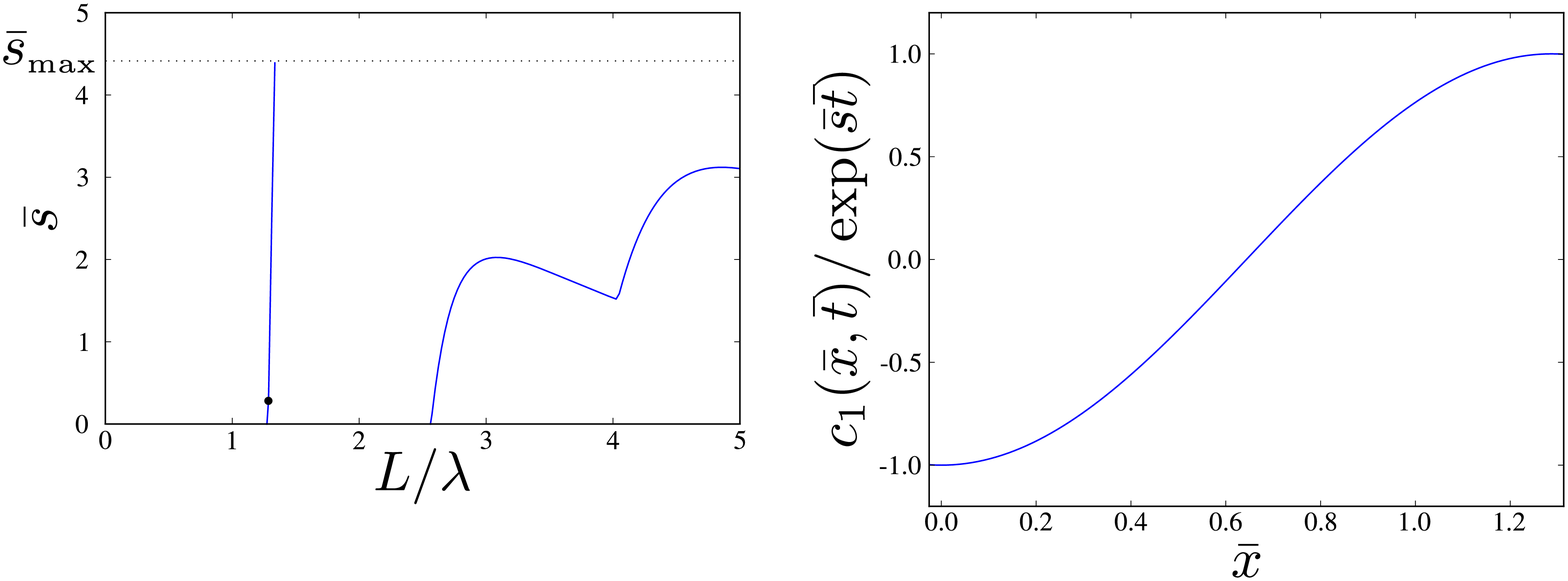}
\end{tabular}
\caption{The cross section of Fig. 1 (b, d) for $c_0=0.5$. a) $\mathcal{R}=0.1$; b) $\mathcal{R}=5$. On the left side the maximal existing dimensionless relative growth rate $\bar{s}$ of noise $c_1$ (vertical axis) via dimensionless length of particle $L/\lambda$ is shown. On the right side we plot the noise function $c_1(\bar{x},\bar{t})/\exp(\bar{s}\bar{t})$ having the following parameters
a) $L/\lambda=0.27$, $\bar{k}_1=1.65$, $\bar{k}_2=1.81$, $\bar{s}=4.46$, $(A_1,A_2,A_3,A_4)=(-0.17,-0.75,0.15,0.62)$,
b) $L/\lambda=1.29$, $\bar{k}_1=0.375$, $\bar{k}_2=2.42$, $\bar{s}=0.41$,  $(A_1,A_2,A_3,A_4)=(-0.085,0.02,0.01,-0.996)$.
$L/\lambda$ is the minimal value where the system is unstable. This point is noted by small black circle on the left plots.}
\end{figure*}

In Fig. 2 we show for a) $\mathcal{R}=0.1$ and b) $\mathcal{R}=5$ the cross section $c_0=0.5$, then for small $L/\lambda$ we shown respective amplitude function $c_1(x,t)/\exp(st)$ and printed respective $\bar{k}_1$, $\bar{k}_2$, $A_1$, $A_2$, $A_3$, $A_4$, so anyone can check the relative growth rate by substituting this solution into linearised equations (10-13). We also find that for small length of particle $L/\lambda$ due to noise $c_1$ we do not have decomposition of $c$ but rather increase or decrease of $c$.

{\bf Conclusion.} 
From figures we can see that analytical result for $\mathcal{R}=0$ is the same as in Fig. 2 of
Ref. \cite{BurchBazant2009}. But for infinitesimal influx rate $\mathcal{R}$ we get maximal spinodal gap, which means instability even for very small particles due to influence of boundary conditions. For size of particle $L$ roughly less than the diffusion length $\lambda$ usually only noise $c_1(x,t)$  having zero $x$ roots will survive (Fig 2.), that is the initial unstable concentration will increase or decrease which will cause some nano particles to lose and others to get more Lithium in the same condition. Then for greater influx rate $\mathcal{R}$ the system will become more and more stable to some limit (look f) in Fig. 1), where we still have small instability stripped region for small $L/\lambda$, which will not disappear for $\mathcal{R}\rightarrow\infty$. Therefore the instability region sometimes cannot be simply described by spinodal gap for enough big $\mathcal{R}$.

We check and prove the results by direct substitution of calculated noise $c_1(\bar{x},\bar{t})$ into linearised equations, and we get correct relative growth rate with numerical error within 0.1\%.

{\bf Acknowledgement}

We thank the german research foundation for financial support under the running DGF Priority Program SPP 1473. MF and HE acknowledge the financial support from the Oberfranken-Stiftung and the state of Bavaria of the Federal Republic of Germany.

\bibliography{reference.bib}

\end{document}